\documentclass[12pt,preprint]{aastex}

\shorttitle{Polarization in Obscured Stars}
\shortauthors{Kuhn et al.}

\begin{document}

\title{A New Mechanism for Polarizing Light from Obscured Stars}

\author{J. R. Kuhn$^1$, S. V. Berdyugina$^2$, D. M. Fluri$^2$,  D. M. Harrington$^1$ and J. O. Stenflo$^2$}
\affil{$^1$Institute for Astronomy, University of Hawaii, Honolulu-HI-96822}

\affil{$^2$Institute of Astronomy, ETH Zurich, CH-8092 Zurich}

\begin{abstract}
Recent spectropolarimetric observations of Herbig AeBe stellar systems
show linear polarization variability with wavelength and epoch near
their obscured H$_\alpha$ emission. Surprisingly, this polarization is
$not$ coincident with the H$_\alpha$ emission peak but is variable near
the absorptive part of the line profile. With a new and novel model we
show here that this is evidence of optical pumping -- anisotropy of the
incident radiation that leads to a linear polarization-dependent
optical depth within the intervening hydrogen wind or disk cloud. This
effect can yield a larger polarization signal than scattering
polarization in these systems.

\end{abstract}


\section{Introduction}

Optical polarization measurements may reveal information
about the spatially unresolved conditions in the near-environment of
stars imbedded in gaseous disks and outflowing winds. Several authors
have advanced agendas for using linear polarization measurements to
infer the structure of the gas and dust surrounding, for example,
Herbig AeBe stars. The analytic models of McLean (1979) used the effects of polarized Thomson scattering and unpolarized line
emission to generate variable linear polarization across line profiles. Wood, Brown, and Fox (1993), 
and Vink et al. (2005) extended this idea in order to predict and model in
detail wavelength variations in the polarization near nebular emission
lines. Their model accounts for  the asymmetry of the nebular scattering and the optical
depth dependence of the scattering source. Harries (2000) has also built
realistic monte-carlo scattering models of hot-star winds. His results are
notable in that they also see absorptive polarization effects that 
may not be due to
the dilution of the scattering polarization from unpolarized line radiation. 

Harrington and Kuhn (2007, henceforth HK) observed a sample of 5 Herbig
AeBe stars that exhibited significant linear polarization variations in
wavelength near the {\it absorptive} part of the H$_\alpha$ line. Their
largest polarization signal, in AB Aur, was about 1\% in the blue
absorptive region of the P-Cygni profile, although if expressed as a
percentage of the continuum intensity, this is about 0.5\%. Even
though their measurements were sensitive to linear polarization at the
level of a few parts in $10^4$, they found negligible polarization variations
at the H$_\alpha$ peak emission wavelength.   The HK results are
problematic for models which depend on line radiation to dilute
polarized Thomson scattering, as these
models tend to predict significant polarization near the H$_\alpha$
peak emission wavelength. This has led us here to consider resonant scattering
models.

Resonant polarized scattering has been considered in the absence of
magnetic fields (Warwick annd Hyder 1965) and in the presence of
depolarizing magnetic fields through a mechanism known as the Hanle
effect (cf.  Hanle 1924, Stenflo 1994) and to describe scattering in
stellar envelopes (cf. Ignace, Nordsieck, Cassinelli 2004). From a
quantum mechanical perspective, resonant scattering polarization is
caused by unequal magnetic substate populations within the upper level
of an atomic transition.  This can be induced by 
anisotropy of the incident radiation and 
correlations between the upper sublevels which yields polarization
of the scattered/re-radiated light (cf.  Stenflo
1998). Resonant scattering alone probably cannot account for the HK observations,
but a resonant $absorption$ process appears feasible.

Polarized absorption results if the lower state magnetic atomic sublevels are
unequally populated, for example by anisotropic incident radiation. Such an ``optically pumped'' gas has an opacity
that will depend on the electric field
direction of the incident flux. Thus, in general, unpolarized light
incident on such an absorbing gas can emerge polarized without scattering.
Optical pumping has been demonstrated in the laboratory (Happer 1972) and has
been discussed in the context of sensitive solar observations (cf.
Trujillo Bueno and Landi Degl'Innocenti 1997).

This mechanism has a simple classical explanation when the lower level
has, for example, total angular momentum $j=1$ 
(which is nominally degenerate with magnetic sublevels $m=\pm$1,
and 0), and the upper level has $j=0.$  In a cartesian coordinate system
we associate the electronic transitions or 
substates with classical electronic 
oscillators aligned in the {\it x-y} plane (for $\Delta m=\pm$1) and in the $z$
direction (for $\Delta m=0$). We impose a 
coordinate system with the pumping radiation incident on the gas in the
$+z$ direction. Then, only the $\Delta m=\pm 1$ transitions can be excited because the incident transverse
electric field lies in the $x-y$ plane.   On the other hand
the subsequent spontaneous downward transitions equally populate all
lower magnetic sublevels. Thus, if there are no collisions to mix
sublevels, eventually only the $m=0$ substate will be populated. This
case has been analyzed for an anisotropic stellar atmosphere
in Trujillo Bueno and Landi Degl'Innocenti (1997).

Now if a second unpolarized beam is incident on the gas in the $+x$
direction (having equal $y$ and $z$ electric field components) only the $z$
component of this electric field is absorbed and scattered by the  pumped gas, because only $m=0$ electronic groundstate oscillators are populated.
Thus the emergent beam is linearly polarized with a $y$-direction
dominant electric field.  In general a gas which is anisotropically
excited, but is observed with light which is incident from a direction
that differs from the optical pumping beam direction, will exhibit an emergent linear
polarization with dominant electric field in a direction which is
perpendicular to the plane of incidence with the pumping beam. This is
the geometry we expect from a star imbedded in a disk or
outflowing wind. 

When the intervening cloud is optically thin the
polarization of the absorbed spectral feature can be larger than any
scattered light spectral polarization signal. Figure 1 shows the
geometry of a gas cloud near a star of luminosity $L$. Let the 
cloud and star have radii $r_c$ and $r_s$ and let the cloud be a distance $d$
from the star. The solid angle subtended by the cloud and the
observer's detector, as seen from the center of the star, are $\Omega_c$
and $\Omega_0$. The relative polarization of the scattered light signal, $Q_{sc}$, to the
total continuum optical signal, $I_s$, when the cloud is not projected against 
the disk of the star, is $$Q_{sc}/I_s = {L\Omega_c/4\pi ~f \Omega_0/4\pi ~p_s\over L\Omega_0/4\pi} ~=~
p_s f {r_c^2\over 4 d^2}$$ where $f$ is the fraction of light incident on the
cloud which is scattered (assumed isotropically) and $p_s$ is the 
average intrinsic polarization of this scattered light. 

On the other hand, when the cloud lies directly between the disk of the
star and the observer, optical pumping can cause the absorbed light to be
polarized. Here {\it all} the light removed from the beam, by scattering
away from the direction to the observer,  appears as a
polarized absorption spectral feature. This contrasts to the off-disk 
scattering geometry where
it is just the light scattered by the cloud into the small solid angle
of the distant detector which is polarized. This absorptive geometry
can lead to a larger polarized optical signal. If $r_c
< r_s$ then the relative transmitted polarized flux, $Q_{tr}$,
normalized to the total, is $$Q_{tr}/I_s = {(r_c/r_s)^2 L \Omega_0/4\pi
~(1-f) p_a\over L\Omega_0/4\pi} = p_a f{r_c^2 \over r_s^2}.$$ Here $p_a$ is
the absorptive (optically pumped) intrinsic polarization which we compute below.  The
intrinsic scattered polarization $p_s$ and $p_a$ are due to population differences 
in, respectively, the upper and lower level magnetic substates, and to the anisotropy of the
scattered or pumping radiation. If these intrinsic
polarization levels are comparable then the ratio of the total
absorptive to scattered polarization is simply ${Q_{tr}/Q_{sc} \approx
4d^2(1-f)/ fr_s^2}$.  Since the cloud lies above the star, $d$ is always
larger than $r_s$ and with $f\approx 0.5$ the absorptive polarization signal is (subject to these assumptions) larger than the scattered polarization
signal.

Our calculation simplifies the general problem, in particular for the case where
the ``cloud'' has a small line-of-sight velocity with respect to the star. 
In this case it is not stellar continuum radiation, but H$_\alpha$ emission
that optically pumps the intervening gas. In general this enhances the
pumping (since H$_\alpha$ can be significantly brighter than continuum) and
complicates the geometry (because of the stars rotational doppler shifted photospheric pumping radiation source). Thus our discussion here applies
directly to the outflowing wind configuration of, for example, AB Aur. The
disk systems HK observed will require a slightly different model.

\section{Statistical Equilibrium and H$_\alpha$}

In order to compute the intrinsic absorptive polarization $p_{a}$ we solve the 
statistical equilibrium equation for the
densities of the first 28 (LS coupled) 
hydrogen energy levels corresponding to all of the 
principal quantum numbers $n=1,2,3$ and $l=0,1,2$ sublevels. Thus the ground state ($n=1$)
consists of an $l=0,~j=1/2,~m=\pm 1/2$ doublet. The $n=2~(l=0,1)$ levels include two $j=1/2$ and one
$j=3/2$ states for a total of 8 magnetic substates. Similarly there are 18 $n=3~(j=1/2,3/2,
~5/2)$ sublevels. A 28$\times$28 matrix statistical equilibrium equation is constructed from the radiative 
transition rates. We note that collisional transitions  are
unimportant in the density regimes of interest here. 
The solution for the individual sublevel densities is obtained by a 
singular value decomposition method.

We first obtain theoretical line strengths and Einstein $A$ and $B$ coefficients for all
allowed dipole transitions between individual sublevels (Sobelman 1992). These coefficients describe fixed
upper and lower magnetic substates so that stimulated emission and absorption
$B$ coefficients for each transition are equal. We check our calculations against the fine structure $A$ coefficents and line strengths listed by NIST
(Ralchenko et al. 2007).
We again take the quantization axis to lie along the mean incident pumping 
radiation direction and take the opening half-coneangle as $\theta$. It is the angle-averaged mean radiation intensity
that multiplies the Einstein $B$ coefficients in the rate equation. We
derive distinct geometrical scaling factors 
for $\Delta m=0$ and $|\Delta m|=1$ 
transitions that follow from integrating Stenflo's (1994) eqns. 3.72 over the pumping
radiation solid angle, $\Omega_s=\pi\theta^2$. For a limb-darkened illuminating
stellar disk where $\mu =\cos\gamma$ and $I(\mu) = I_0(1-a+a\mu)$ we obtain factors $C_{|\Delta m|}(\theta)$
that multiply the pumping blackbody background radiation terms in the rate equation. We
find $C_0(\theta ) = (1-a)(1/2-3\cos{\theta}/4+\cos^3{\theta}/4)~+~a(3/8(1-\cos^2{\theta})+3\cos^4{\theta}/16)$ and 
$C_1(\theta )=(1-a)(1/2-3\cos{\theta}/8-\cos^3{\theta}/8)~+~a(9/16-3\cos^2{\theta}/16-3\cos^4{\theta}/32)$. We use the theoretical limb-darkening coefficients from Al-Naimiy (1978).

Anisotropy of the illuminating stellar light source leads to a difference in
the $m=\pm 3/2$ and $m=\pm 1/2$ populations (within the $n=2, ~j=3/2$ level). 
This lower state anisotropy linearly polarizes the transmitted
light from the star which
passes through the cloud at a non-zero angle, $\gamma$, with respect to the
$z$ (optical pumping) direction. 
Given the geometry defined by Fig. 1 we expect the emergent 
light to be partially polarized with electric field perpendicular to the plane of the 
figure. Most astronomical observations do not resolve the fine structure of
the $n=2$ to $n=3$ H$_\alpha$ line but they will still exhibit a diluted linear polarization signal 
because only some of the transitions are insensitive to the 
pumping radiation anisotropy (e.g. those which start and end on $j=1/2$ levels).

\section{Polarized Radiative Transfer in the Intervening Cloud and the Form
of the Polarization Solutions}

The Stokes transfer equation simplifies in our geometry where the stellar photosphere
is a background light source and the intervening cloud is optically thin. We take the
positive Stokes $Q$ direction to correspond to light with electric field perpendicular to the 
plane of Figure 1. The Stokes vector transfer equation (cf. eq. 11.1
Stenflo 1994) involves only $I$ and $Q$ intensity, and $\eta_I$ and $\eta_Q$ continuum-normalized opacity
terms. Since the stellar photospheric background 
is the only source of photons observed through the cloud,
there are no source or emissivity terms to consider in the polarized transfer
equation. 

Harrington and Kuhn (2007) detected only small polarization ($Q << I$) in the stellar systems they observed. 
In this approximation, the transfer equation reveals that
light transmitted through a diameter of the cloud will have polarization 
$p_{a}=Q/I=\Delta Q/I~\approx 2r_c\eta_Q\kappa_{cont}$ where $\kappa_{cont}$ is
the normalizing continuum opacity. Here $I$ is the continuum intensity
and we compute $\eta_{I,Q}\propto n_{I,Q}$ from the lower level population anisotropies and 
eqns. 3.74 (Stenflo 1994). 
In this case the effective refractive index and opacity terms $n_{0,\pm}$ or H$_{0,\pm}$ 
(eqn 4.41) are computed from sums of the products of the $B$ coefficients and the sublevel populations, derived separately for the 
$\Delta m=\pm 1$ and $\Delta m=0$ transitions. The
optical depth through the cloud is just $\tau = 2r_c \eta_I\kappa_{cont} \approx -\log{(1-f)}$ where $f$ is 
the fraction of H$_\alpha$ photons scattered, as defined above. We find that for
clouds with typical path lengths of, say, $ r_s$ the optically thin condition
yields hydrogen densities that are effectively collisionless.


In the low density conditions of this disk/wind the statistical equilibrium
rate equation depends on only stimulated emission, absorption, and
spontaneous emission processes.  It is the difference in the angular
dependence of the $\Delta m = 0$ versus $|\Delta m | = 1$ coupling to
the anisotropic radiation field that leads to differences in the
degenerate $n=2$ magnetic sublevel populations. Several
qualitative features of the optical pumping solutions are intuitive.
For intense pumping radiation fields the anisotropy yields a decreasing
absorptive polarization -- in this regime the spontaneous emission is
negligible and the net upward and downward transition rates between any
two levels must be the same. In this case the equilibrium level populations
are independent of the upward/downward transition rates. Consequently there can be no  sublevel density
differences and the absorptive polarization must vanish. Similarly
for weak radiation fields it is the spontaneous emission that defines
transition rates and this term in the rate equation does not depend on
the radiation anisotropy.  Thus at low and high radiation backgrounds
the absorptive polarization approaches zero.

The absorptive polarization near H$_\alpha$ is also a function of the ratio of the
H$_\alpha$ to $L_\beta$ (and $L_\alpha$) pumping flux. The ratio of
the $n=2,~j=3/2$ to $j=1/2$ populations affects the net H$_\alpha$
polarization, since
only the $j=3/2$ states are sensitive to radiation anisotropy. In the limit of no 
background H$_\alpha$ flux (and in the absence of collisions) all the hydrogen ends up in the ``sterile''
$n=2~l=0$ state because it has no decay route and there is no radiative absorption to depopulate it.  
Conversely, as the pumping H$_\alpha$ flux increases the equilibrium $n=2~j=3/2$ levels are populated and the induced
linear polarization can increase.

This results in a rather weak temperature dependence of the optical pumping source on the
absorptive polarization. Even though a decreasing source temperature decreases the
pumping radiation, this effect on the polarization is compensated by the 
increasing ratio of H$_\alpha$ 
to $L_\beta$ flux. We find that between 30,000 to 3,000K, a change in pumping
flux by nearly 4 orders of magnitude, the absorptive polarization is nearly constant.
For example, a cloud located 1.1 stellar radii from a 30,000K star
($\theta = 65\degr$) that is $\gamma =65\degr$ from the line-of-sight to the center of the
star, yields linear polarization of 0.53\%. If the star has a temperature of 3000K the
polarization only decreases to 0.48\%.

For fixed radiation mean intensity, as the pumping half-cone angle ($\theta$) increases
the anisotropy decreases and the polarization decreases. This effect continues until
$\theta =90\degr$, at which point there is no difference between the dipole coupling of radiation
to $\Delta m=0$ and $|\Delta m|=1$ transitions. The actual level population difference
scales with $\theta$ as the difference between the derived geometric radiation coupling
coefficients $C_0(\theta )$ and $C_1 (\theta )$ defined above. It is straightforward to
see that this varies like $\theta^2$ at small angles. Thus the linear polarization $decreases$ for small cone angles (even though the anisotropy increases). This is because the pumping radiation intensity declines
rapidly with $\sin{\theta}=1/d$ (where $d$ is in units of the stellar radius) as the
cloud-star distance increases. Figure 2 shows how the linear polarization varies for a
cloud seen at the projected edge of a star, as a function of the cloud-star distance, $d$. At small distances from the star the limb darkening function
yields significant anisotropy that keeps the polarization from going to zero.

The absorptive polarization depends strongly on the angle between the absorbed beam and the
mean pumping direction. As this angle ($\gamma$) increases the polarization perpendicular to the plane increases. For example, with a $T=10,000K$ source
temperature and a $65\degr$ cone angle the polarization increases almost linearly from $0$ to $0.5\%$ as $\gamma$  varies from $0$ to $65\degr$.

\section{Specific Solutions}

We are developing detailed forward solutions for all the HAeBe stars we have observed and to account for stellar emission line pumping. Here it is interesting to describe the general form of the wavelength dependent linear polarization
expected from an intervening wind or disk structure. We consider two geometries; a radial
outflow and a rotating disk illuminated (optically pumped) by the stellar blackbody. The
geometry is essentially that defined in Fig. 1 but we allow that the wind or disk's symmetry
axis is inclined at an angle $b$ with respect to the line-of-sight to the star.
We rotate the common plane containing the
line-of-sight direction and the symmetry axis to be oriented vertically. Then, as seen from
the observer, the intervening cloud projects onto a horizontally oriented arc 
across the stellar disk. The disk is both
illuminated by the star and absorbs the stellar light behind it. Each point on this arc
contributes absorptive polarization at a distinct velocity with respect to the observer.

We take positive Stokes $Q$ along the vertical sky direction.  Positive $U$ 
corresponds to an electric field that is $45\degr$ to this. It follows that the absorptive polarization of a symmetric radial wind should yield only $+Q$ polarization at all velocities. In Figure 3 we've plotted the $Q$ polarization across
the velocity distribution of the intervening outflow in units of the maximum (assumed spatially
constant) outflow velocity. This is illustrative, although  most real stellar disks and outflows are not expected to be homogeneous in density or velocity.
In general we expect (and observe) a non-zero $U$ polarization which, of course, depends on the rotation angle between the instrument $Q$-axis and the star. The
zero polarization level is also ill-defined in the observations since a
wavelength independent constant has been removed. Nevertheless the polarization
amplitude and simple form of AB Aur in the HK data is consistent with the ``diamond" symbols in Fig. 3. 

In contrast, even a homogeneous rotating disk geometry must exhibit non-zero $Q$ and $U$ line profiles because the
zero line-of-sight velocity condition occurs along the vertical symmetry line
near the peak of the H$_\alpha$ emission. The characteristic symmetric (about zero relative velocity) Stokes
$Q$, and antisymmetric $U$ profiles for such a disk are also plotted in Figure 3.  These are plotted against units of the maximum rotation or outflow velocity of the disk or wind.
The polarization outside of these plotted points  must be
zero, but spectral smoothing due to the finite spectral resolution of the measurements will yield some polarization outside of the absorptive region.

\section{Conclusions}

This new model shows how optical pumping can describe the magnitude and form of the absorptive linear polarization
features recently seen in some obscured stellar systems. The variation of $Q$ and $U$ linear
polarization detected near H$_\alpha$
originates from the intervening gas cloud  projected against the disk of the star. The polarization features contain 
information on the kinematics and inclination of the cloud. One immediate
conclusion from these calculations is that the absorptive region is near the
surface of the star, most likely within 2 stellar radii in order to have
a polarization amplitude of about 0.5\% . For constrained geometries it may
be possible to devise an inversion technique to extract more specific
information on non-homogeneous cloud structures in the innermost regions
of these stellar systems.

\acknowledgements
This research was supported by the NSF through grant AST0123390. SVB
acknowledges the EURYI Award from the ESF. JRK is
grateful for local support from the ETH/Zurich, where much of this work was completed, and to Don Mickey, for helpful comments on the manuscript.

\clearpage
\begin{figure}
\plotone{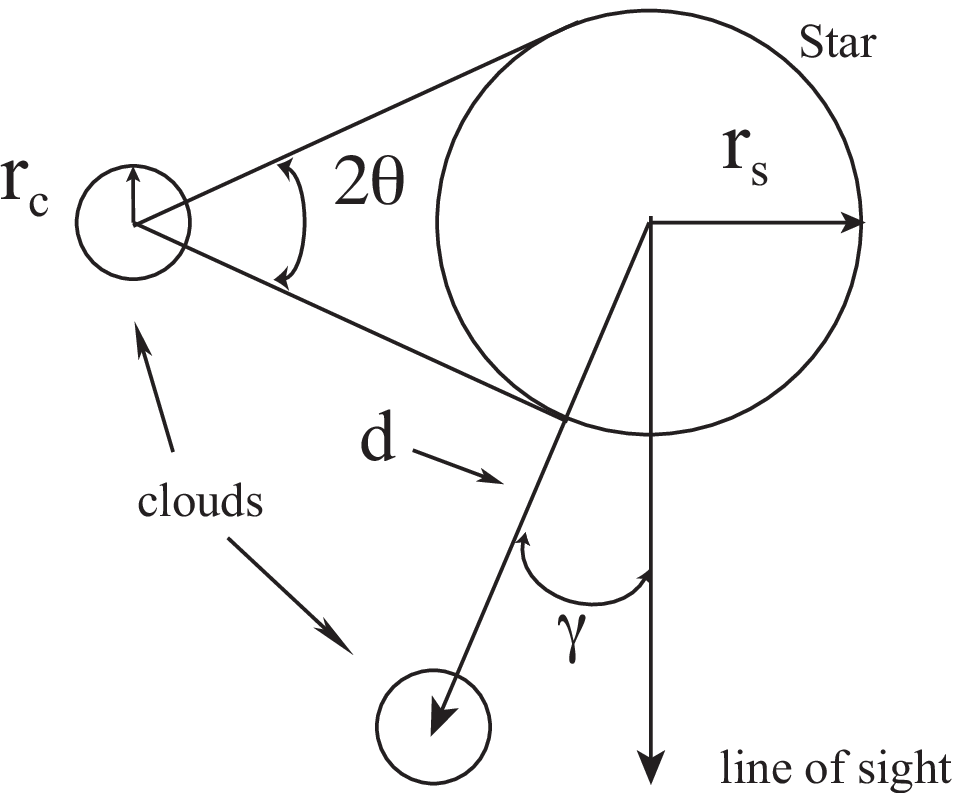}
\caption[jfig2.eps]{\label{fig:fig1}
Off-limb and intervening gas cloud geometries that yield H$_\alpha$ polarization}
\end{figure}

\clearpage
\begin{figure}
\plotone{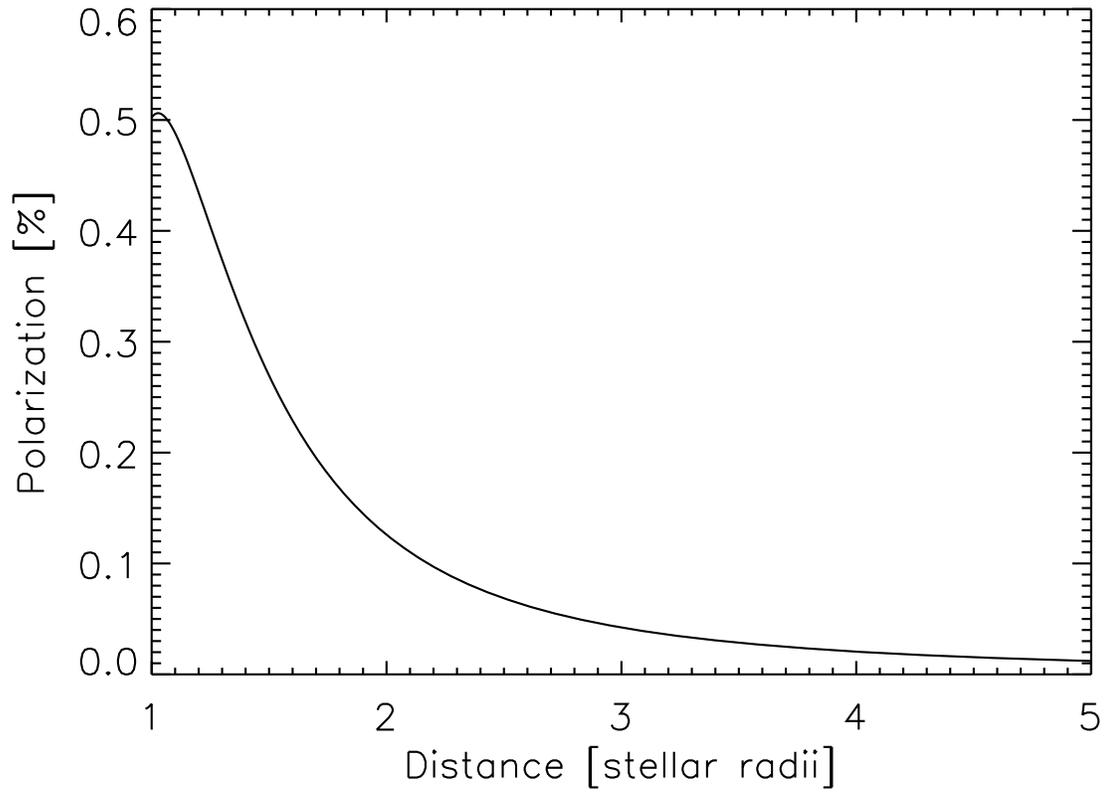}
\caption[optpumpfig2.ps]{\label{fig:cmd}
Polarization as a function of distance, $d$, between cloud and star assuming a T=10,000K A star, limb darkening, and with the intervening
cloud at the projected edge of
the stellar disk at the indicated radial distance from the star center.}
\end{figure}

\clearpage
\begin{figure}
\plotone{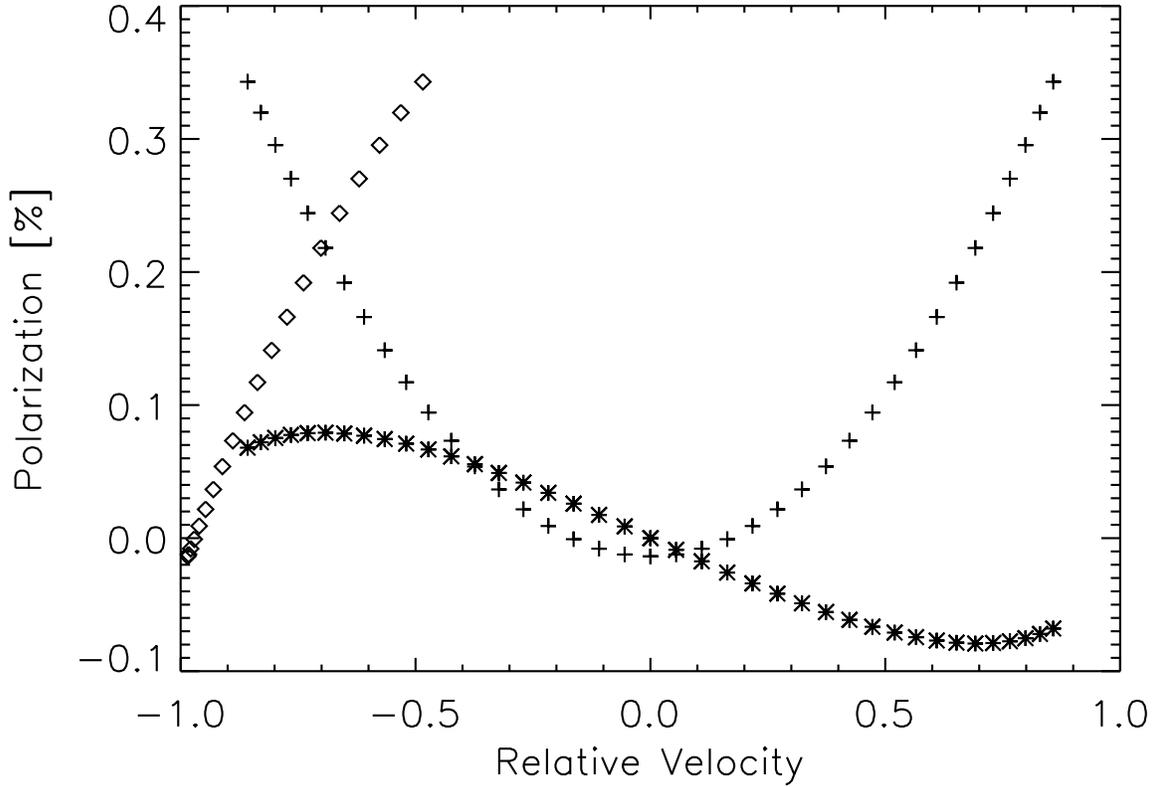}
\caption[optpumpfig3.ps]{\label{fig:cmd}
Stokes $Q/I$ (plus symbols) and $U/I$ (star symbols) polarization versus relative velocity of a 
rotating hydrogen disk. The disk inclination angle is $b=80^o$, $T=10000$K, and $d=1.1 r_s$. The
$Q/I$ polarization for a comparable outflowing wind is plotted with ``diamond'' symbols against
units of the relative maximum outflow velocity. The peak H$_\alpha$ emission
occurs at zero in the horizontal axis units and the polarization signal in each case is zero outside of the plotted points for that geometry.  }
\end{figure}

\end{document}